\documentclass[sigconf]{acmart}

\begin{document}

\title{Unveiling Coordinated Groups Behind White Helmets Movement}
\title[Unveiling Coordinated Groups Behind White Helmets Disinformation]{Unveiling Coordinated Groups Behind White Helmets Disinformation}

\author{Diogo Pacheco, Alessandro Flammini, Filippo Menczer}
\email{pacheco@iu.edu}
\affiliation{%
  \institution{Center for Complex Networks and Systems Research \\
Luddy School of Informatics, Computing, and Engineering \\
Indiana University, Bloomington}
}


\begin{abstract}
Propaganda, disinformation, manipulation, and polarization are the modern illnesses of a society increasingly dependent on social media as a source of news.
In this paper, we explore the disinformation campaign, sponsored by Russia and allies, against the Syria Civil Defense (a.k.a. the White Helmets). 
We unveil coordinated groups using automatic retweets and content duplication to promote narratives and/or accounts.
The results also reveal distinct promoting strategies, ranging from the small groups sharing the exact same text repeatedly, to complex ``news website factories'' where dozens of accounts synchronously spread the same news from multiple sites.
\end{abstract}


\acmConference[WWW '20 Companion]{Companion Proceedings of the Web Conference 2020}{April 20--24, 2020}{Taipei, Taiwan}
\acmPrice{}
\acmDOI{10.1145/3366424.3385775}
\acmISBN{978-1-4503-7024-0/20/04}


\maketitle

\section{Introduction}

Television is only the main source of news for older Americans as younger generations are shifting towards the web~\cite{shearer2018social}. 
And social media has been playing a major role shaping what news are seen and when. 
Paradoxically, despite having access to more resources and information than ever, people are becoming more vulnerable to misinformation. 
Information overload, finite attention, social pressure towards being updated (knowing the trends), 
and
the lower cost of sharing opinions 
are creating
a fertile environment for propaganda, disinformation, and manipulation campaigns~\cite{bessi2016social,ferrara2017disinformation, Lazer-fake-news-2018,vosoughi2018spread,stella2018bots,Shao2018,deb2019perils,Bovet2019,Grinberg2019}.

Although propaganda, misinformation, and influence campaigns have always existed~\cite{Jowett2018}, we have been witnessing a considerable increase in state-sponsored cyber warfare~\cite{prier2017commanding} --- 
sometimes internally via censorship and propaganda (e.g., China~\cite{bamman2012censorship}), sometimes across borders by promoting polarization and disinformation (e.g., Russia in the 2016 US elections~\cite{Bovet2019,Grinberg2019}).

Despite steady progress by researchers, journalists, and social media platforms in detecting and combating misinformation and inauthentic accounts, malicious actors continue to operate. 
The attacks keep evolving in an arms race, effectively deceiving the public and amplifying misinformation~\cite{barrett2019,keller2019political}.

The Syria Civil Defence, commonly known as \emph{White Helmets} (WH), was formed in 2014 during the Syrian civil war.
It is composed of volunteers who help with search-and-rescue operations in dangerous areas or in response to bombing. They wear cameras on their helmets recording their missions.
The WH published video compilations to denounce atrocities, violations of human rights, chemical attacks, and targeting of civilians commanded by Syrian President Bashar al-Assad and Russia.
A massive disinformation campaign was mounted in response to discredit the organization through fabricated news claiming, for example, that the WH were terrorists and were trafficking organs~\cite{palma2016snopes}.

The Guardian was one of the first news sources to report \textit{how} the WH were being attacked by a disinformation campaign supported by Russia~\cite{solon2017syria}. The report showed some of the mechanisms used to flood social media with fabricated narratives that would be replicated in several websites to create critical mass against the WH.
Starbird et al.~\cite{Starbird2019} presented a case study related to the White Helmets disinformation campaign.
They gained insights from exploratory analysis of digital trace data. For instance, they performed qualitative analysis of tweets sharing links to a network of suspicious websites. 
They described this campaign as being mostly driven by activists rather than automation. Levinger~\cite{levinger2018master} claims this campaign gained credibility because it was well crafted around a master narrative of decline and re-birth of Russia.

In this paper, we explore two suspicious scenarios in which accounts could be promoting narratives and/or accounts: (i) a simple approach based on fast retweets, and (ii) a more sophisticated one based on similar text. 
Employing the coordination network detection framework~\cite{pacheco2020uncovering}, we are able to find suspicious groups influencing online White Helmets conversations according to both scenarios.
The text similarity coordination network reveals groups with different operation strategies, varying from small groups sharing the same content, to large groups wrapping the same claims in multiple websites.  

\section{Detecting Coordination}

Let us apply the approach described by Pacheco et al.~\cite{pacheco2020uncovering}  to identify coordinated accounts  in the context of White Helmets disinformation on Twitter. In particular, we are interested in accounts that employ some degree of automation and for which the coordination is not disclosed or transparent. Rather than evaluating the maliciousness of such accounts, we wish to unveil attempts to masquerade the actions of a single entity as those of many independent users. Coordinated accounts are often challenging to identify, especially when considered in isolation. Their behavior appears suspicious mostly because multiple accounts are behaving similarly.
We show how different filtering strategies applied to the data can highlight different suspicious behaviors. These must ultimately be inspected by humans to decide on their potentially offensive nature.

\subsection{Dataset}

We use an anonymized dataset provided by the DARPA SocialSim project\footnote{\url{www.darpa.mil/program/computational-simulation-of-online-social-behavior}}; they used GNIP to query English and Arabic terms related to the White Helmets between April 2018 and March 2019. In total, the dataset contains over a million tweets by approximately 42 thousand accounts. The Twitter snippets from posts and profiles presented in the results where obtained by querying the Twitter web interface.



\subsection{Rapid Retweet Network}

\begin{figure*}[ht!]
    \centering
    \includegraphics[width=.90\textwidth]{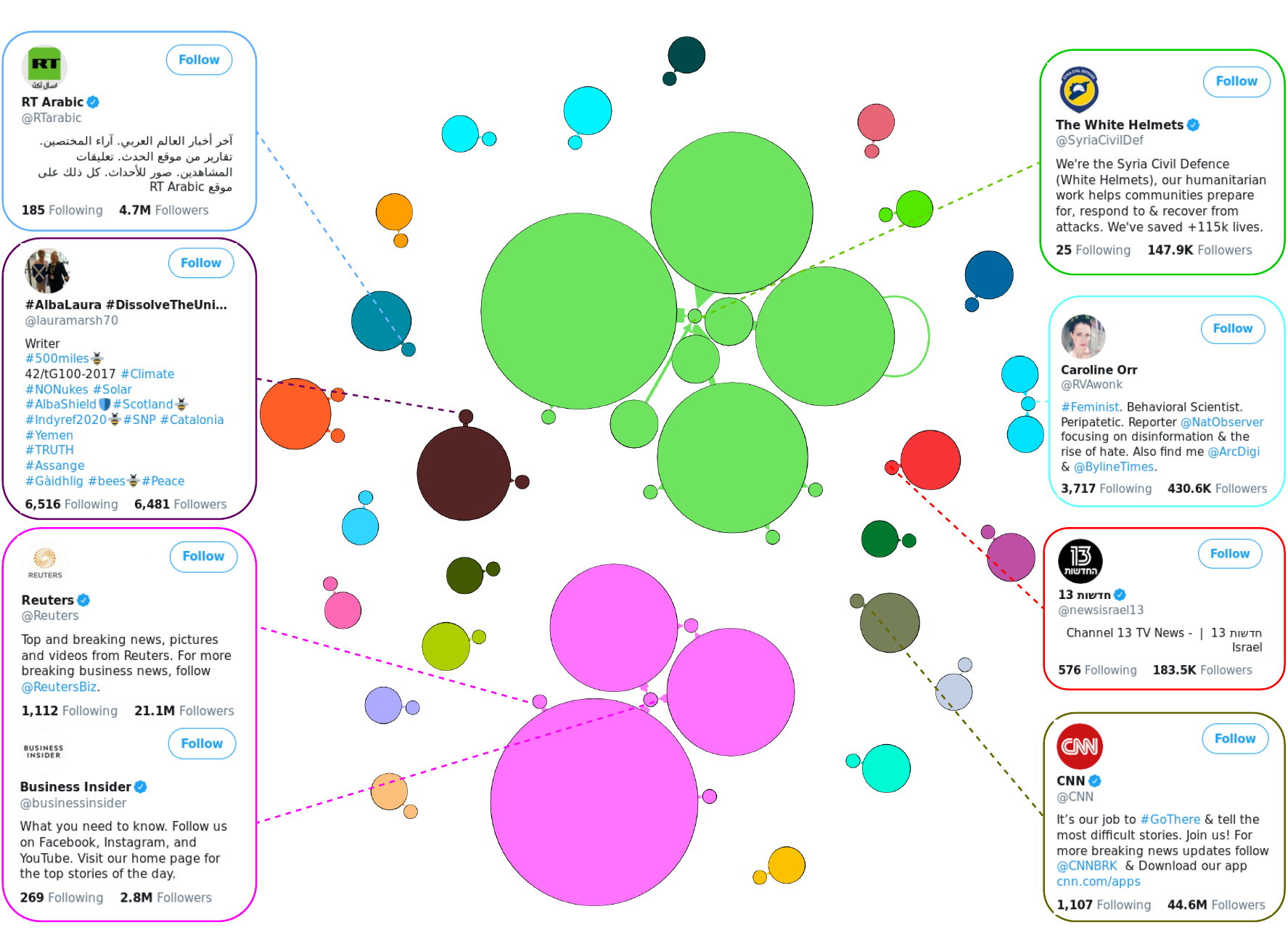}
    \caption{Directed retweet network connecting the retweeter account to the original author. The network is filtered to disregard spurious edges: promoter nodes are only present if they have retweeted the promoted at least twice, and each retweet happened less than ten seconds after the original post. Size is proportional to the number of rapid retweets (out-strength). A few of the accounts being promoted by retweets are highlighted.}
    \label{fig:fast_retweet}
\end{figure*}

\begin{figure}
    \centering
    \includegraphics[width=.95\columnwidth]{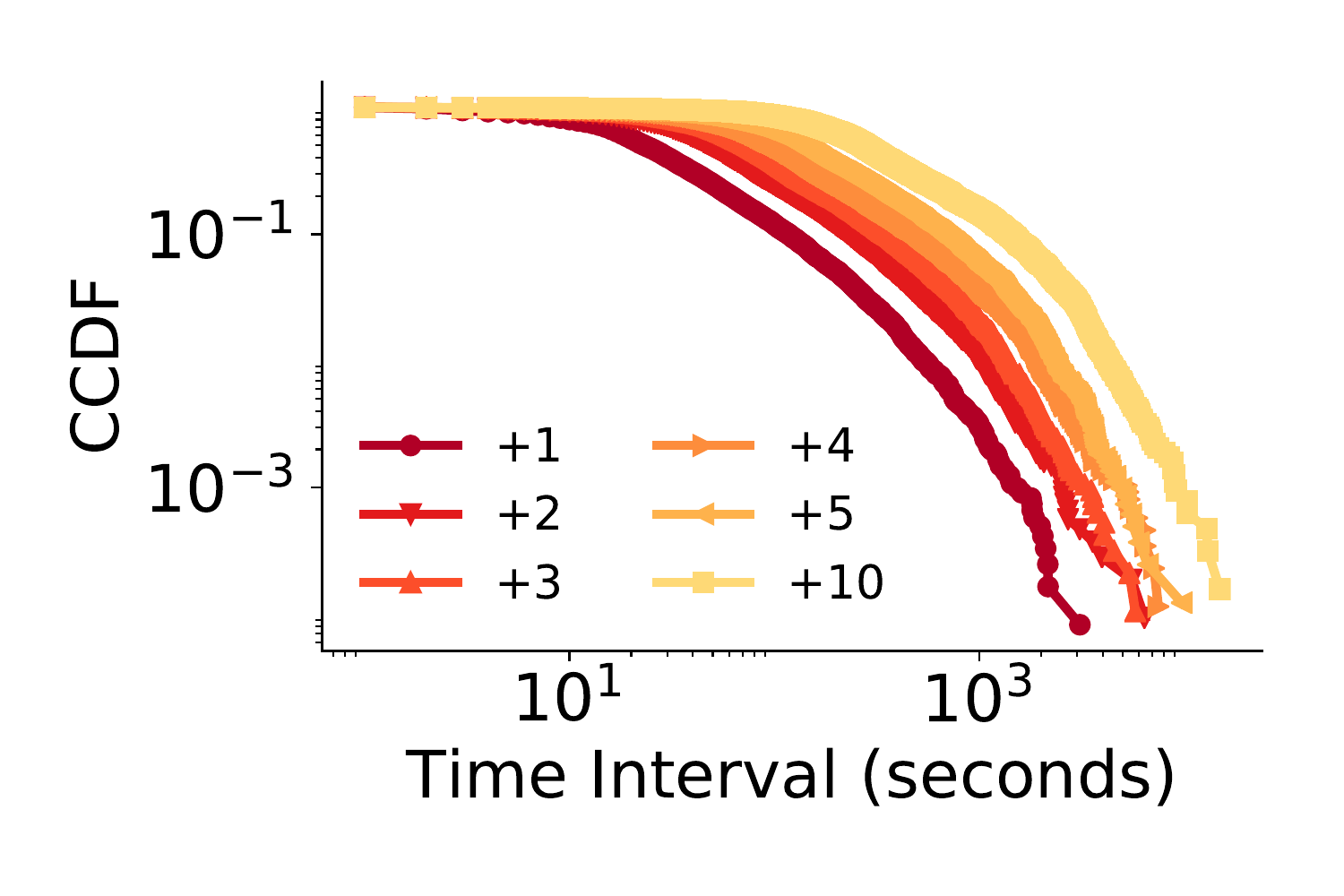}
    \caption{Distributions of time intervals between pairs of tweets with text similarity greater than 0.7. Most tweets replicating content are created a few seconds after the original. The distributions show similar patterns regardless of the sequential distance used to select tweets for comparison.}
    \label{fig:text_similarity_time_interval}
\end{figure}

A first, simplest approach to detect coordination is to identify groups of accounts who consistently retweet the same source. 
We create a direct network by drawing a weighted link from the retweeter (promoter) to the account who produces the original post (promoted). 
While retweeting is the simplest available mechanism to help spread information, Twitter discourages the systematic retweeting of posts from any given source.\footnote{\url{help.twitter.com/en/rules-and-policies/twitter-automation}}
Since retweeting is commonplace, and to avoid labeling chance relationships as suspicious, we only keep edges if the retweet happens within ten seconds after the original post, and if the promoter retweets the promoted at least twice.
Thus, the edge weights correspond to the number of rapid retweets.
Finally, we disregard singletons and extract the connected components of the network to identify  coordinated groups of promoter and promoted accounts.

The resulting network is shown in Fig.~\ref{fig:fast_retweet}. Most connected components are dyads and triads. A node's size is proportional to its out-strength, i.e., to the intensity of its promotional activity. The promoted accounts (smaller nodes) are not engaging in retweeting activity and are usually accounts with many followers. Among them we find several verified and popular media accounts from different countries, such as \textit{RT Arabic}, \textit{Reuters}, and \textit{Channel 13} (from Israel). 

The most evident feature from this analysis is the presence of accounts that systematically retweet popular news sites with great immediacy. 
Although the relationship \textit{promoter}-\textit{promoted} is clearly identified, it provides insufficient evidence to suspect a relation of control of the promoted over the promoter. Indeed, the reputation of the promoted accounts, and the presence of promoters that retweet unrelated news sites indiscriminately --- like in the green or purple clusters in Fig.~\ref{fig:fast_retweet} --- leads to the hypothesis that these are simply automated accounts that independently try to exploit the popularity of the content produced by news sites for their own sake, for example to accumulate followers and gain influence.

\subsection{Similar Tweet Network}

When hunting for suspicious behavior, considering text similarity among original tweets may be more revealing than simple retweeting patterns. 
Replication can have different motivations, such as plagiarism and brokerage. Unlike retweets, tweets with similar content are considered original by the platform, and the links between original and copy are hard to detect.

We adopted a strategy based on text similarity.  
Aiming to identify groups of accounts that deliberately post similar content, we analyzed the text from original tweets, replies, and quotes but ignored retweets. 
We measured text similarity using the Ratcliff/Obershelp Pattern Recognition algorithm~\cite{ratcliff1988pattern}. We considered all pairs of tweets with a similarity above a fixed threshold. We finally built a similar-tweet coordination network of accounts where an edge indicates that the two connected accounts were responsible for at least one pair of similar tweets produced within a short time interval.
The network has two parameters: the text similarity threshold and the time window threshold. Next we determine both empirically.

To reduce the computational complexity implicit in measuring the similarity between all tweets in our database, we first considered the longitudinal stream of tweets and measured text similarities only between tweets separated by at the most nine other tweets along the stream. We sorted all tweets chronologically and used Python SequenceMatcher~\cite{ratcliff1988pattern} to compare the text of consecutive tweets at distance +1 (immediate neighbors), +2 (one tweet in between), and so on, up to +10 (the tenth following tweet).
Fig.~\ref{fig:text_similarity_time_interval} shows that a considerable amount of similar content is created in the first ten seconds after the original tweet post. Therefore, we use this time window threshold to limit the pairwise comparisons between tweets in our collection. 

\begin{figure}
    \centering
    \includegraphics[width=\columnwidth]{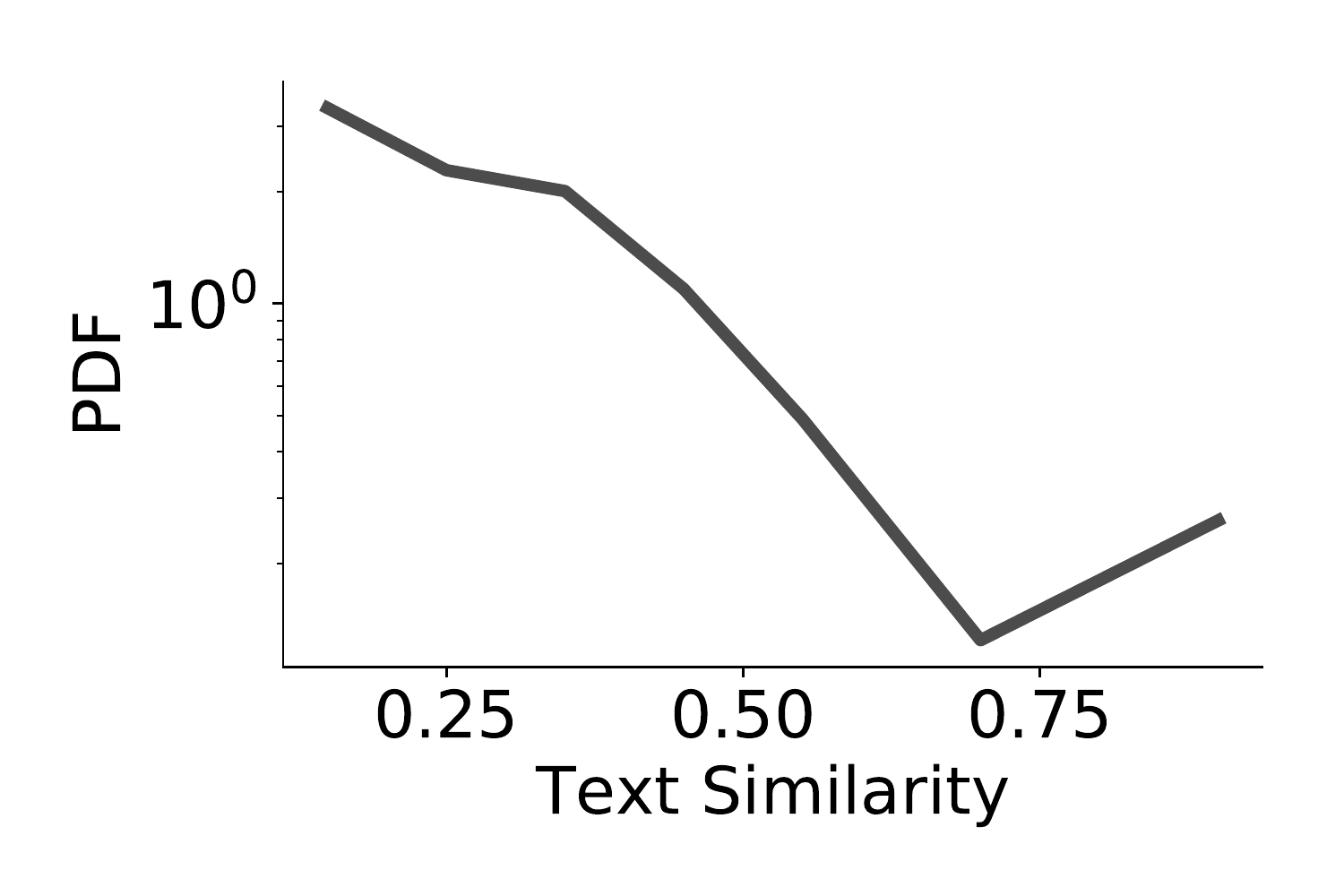}
    \caption{Distribution of text similarity. Pairs of tweets are compared if they were posted less than ten seconds apart.}
    \label{fig:text_similarity_dist}
\end{figure}

Fig.~\ref{fig:text_similarity_dist} shows the distributions of the pairwise text similarity for all pairs of tweets created less than ten seconds apart.
As expected, most content is dissimilar and the probability of finding tweets with increasingly similar content decreases monotonically. A surprising exception is that the probability 
increases for similarity greater than 70\%. Therefore, we set the similarity threshold to 0.7.
%
From now on, for the sake of readability, when we refer to \textit{similar tweets} we mean tweet pairs that have both \textit{content} above the similarity threshold and \textit{creation times} within the time window threshold.


\begin{figure*}
    \centering
    \includegraphics[width=\textwidth]{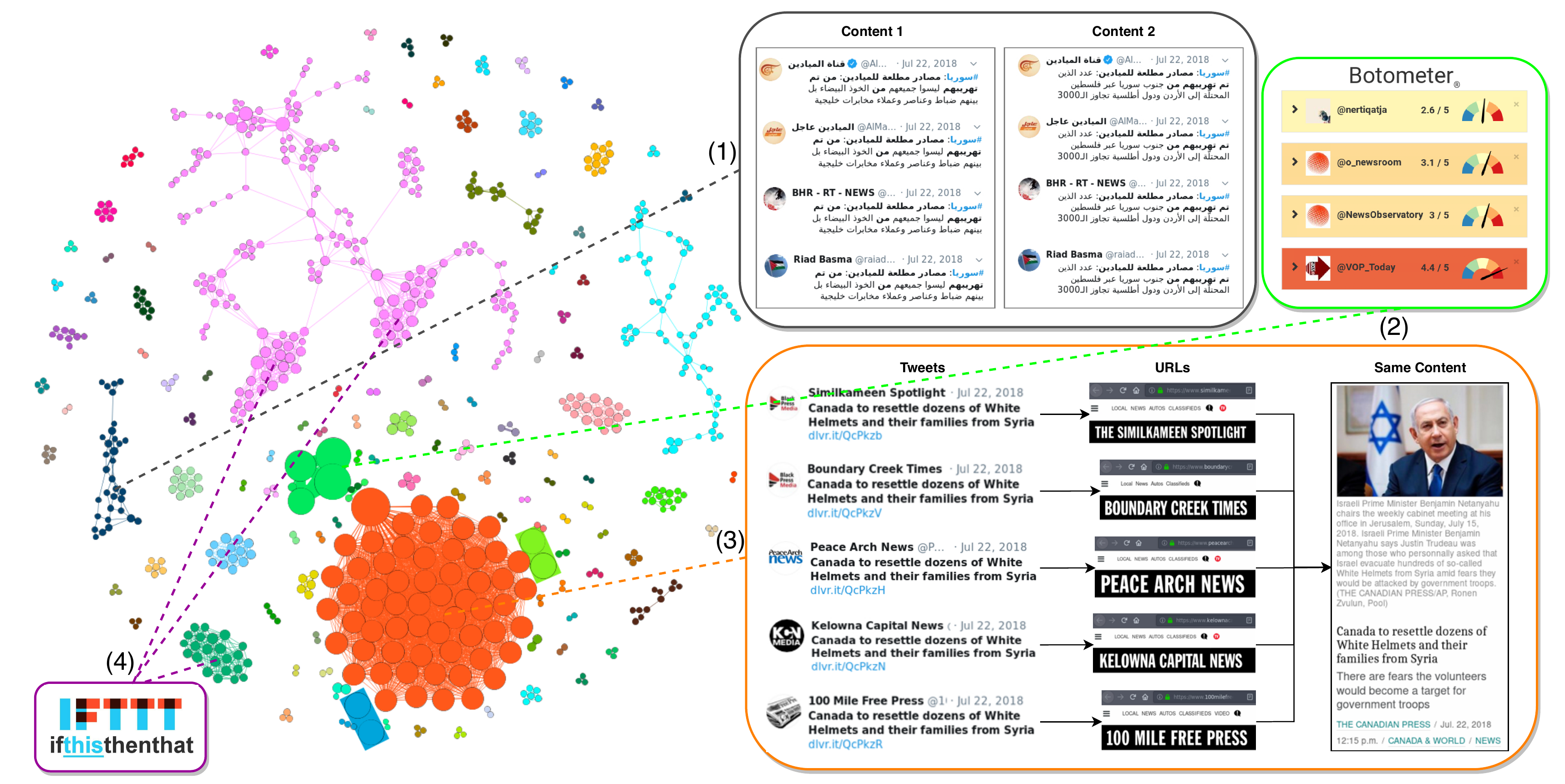}
    \caption{Similar-tweet coordination network. Nodes are Twitter accounts colored by connected components. Two accounts are connected if they tweet text with similarity above 0.7  within 10 seconds of each other. Edge weights represent numbers of similar tweets. Node size is proportional to strength. Singletons and nodes with strength lower than two were filtered out. Four strategies are displayed by highlighted components.  
    (1) Multiple accounts sharing the same content. 
    (2) Few hyper-active accounts with similar tweets, likely to be automated according to Botometer~\cite{Yang2019botometer}. 
    (3) Factory of news websites, with multiple accounts pretending to post news from distinct websites that actually present the same content.
    (4) Multiple groups of accounts using automation services (see Fig.~\ref{fig:ny_example} for an example).
    }
    \label{fig:similar_tweet_net}
\end{figure*}

\begin{figure}[ht]
    \centering
    \includegraphics[trim = 5mm 40mm 5mm 40mm, clip, width=0.9\columnwidth]{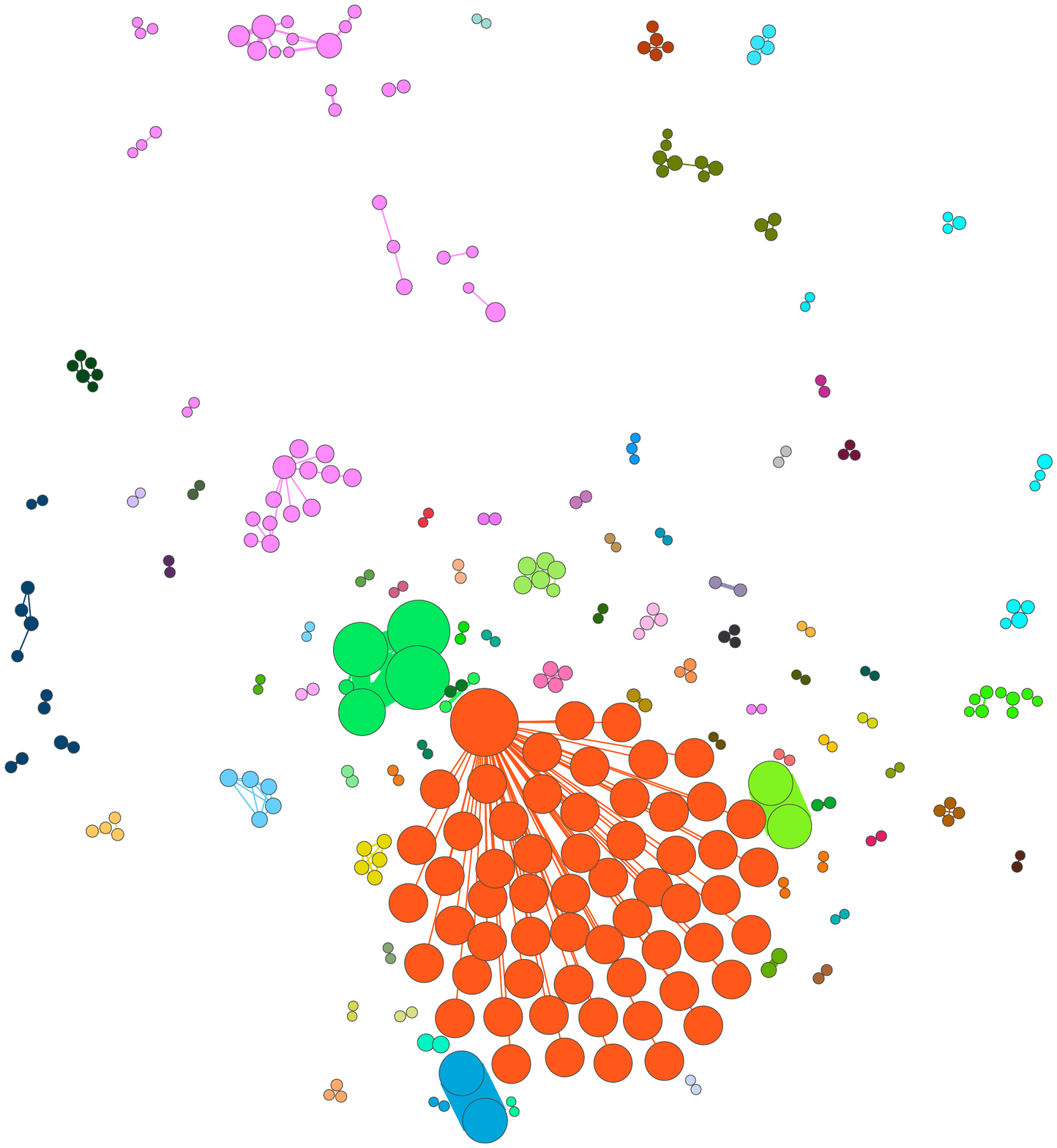}
    \caption{Coordination network as described in Fig.~\ref{fig:similar_tweet_net}, but with a more restrictive minimum edge weight of 2.}
    \label{fig:similar_tweet_net-complement}
\end{figure}

The similar-tweet coordination network is shown in Fig.~\ref{fig:similar_tweet_net}. Two accounts (nodes) are connected if they have posted similar tweets. The weight of a connection represents the number of similar tweets posted by the two accounts. 
To reduce the number of false positives, we apply additional filters.
An account is kept in the suspicious network only if its strength is 
at least two. 
That is, (i) it has a similar tweet to at least two other accounts, e.g., three accounts posting similar tweets; or (ii) it has at least two similar tweets to any one account, e.g., a pair of accounts with multiple similar tweets.

To demonstrate the strategies and behaviors of suspicious accounts, a few coordinated groups are highlighted in Fig.~\ref{fig:similar_tweet_net}:

\begin{enumerate}
    \item \textit{Multiple accounts sharing the same content}---these accounts post several similar tweets, however, not all group accounts are active at the same time, nor is the combination of active accounts the same. Therefore, the group is not a clique.
    
    \item \textit{Few accounts with similar tweets}---a small group of accounts always posting similar tweets. The automation rules and the coordinated behavior of this type of group are more trivial. Thus, current bot detection tools, such as Botometer from Indiana University~\cite{Yang2019botometer}, are able to capture the suspicious behaviors at the individual level.
    
    \item \textit{Factory of news websites}---several accounts sharing a headline and a shortened URL by social media automation tools (often \url{dlvr.it}). In this case, the URLs redirect to several distinct domains hosting allegedly distinct media outlets. However, all these websites look exactly the same (content and style), except for the URL and the name in the top banner.
    
    \item \textit{Accounts using automation services}---several accounts sharing the same headline and the same URL. The URLs are shortened through automation services, such as ``If This Then That'' (IFTTT\footnote{``If This Then That, also known as IFTTT, is a free web-based service to create chains of simple conditional statements, called applets. An applet is triggered by changes that occur within other web services such as Gmail, Facebook, Telegram, Instagram, or Pinterest.''~\cite{ifttt}}), suggesting the posts were also automated. The URLs redirect to well-known news websites such as \emph{The New York Times}, or to intermediate blogs 
    linking 
    to the prestigious websites. Even though some groups are sharing exactly the same content (headline + URL), our method breaks them apart due to longer time intervals between bursts of activity.
\end{enumerate}

We also explored more restrictive filters, for instance, a minimum edge weight of two. That is, two accounts are only connected if they had at least two similar-tweet pairs. This scenario shrinks the coordination network by 60\% (Fig.~\ref{fig:similar_tweet_net-complement}).
Some groups disappear completely, other components are broken up. For example, the largest (pink) component breaks up into multiple smaller groups. Although this approach might eliminate some 
false positives, 
it might miss 
more sophisticated groups in which accounts alternate activation in distinct campaigns. 


\section{Discussion}

The methods used here are not designed to classify coordination as benign/malicious, nor organic/automated. They aim to capture unexpected relationships. Moreover, they shed light on behaviors that are only revealed as suspicious at the aggregate level. For instance, an account sharing a headline promoting a website is ordinary, but a dozen ``different'' websites sharing the same headline with the same content at the same time are not.

Although additional filters can reduce the chances of false positives, it is possible that the method still mislabels a few accounts.
Therefore, careful inspection of the suspicious groups is strongly recommended. 
For instance, a potential source of false positives for the similar-tweet approach is the \textit{share button} present in most news websites. 
When people use this feature to share news on their social media without making any changes in the suggested text, they will end up sharing exactly the same content as independent original tweets. 
If a website is popular, there is a good chance that some of its readers share the same article within a few seconds, creating a suspicious group among them.

\begin{figure*}[t!]
    \centering
    \includegraphics[width=\textwidth]{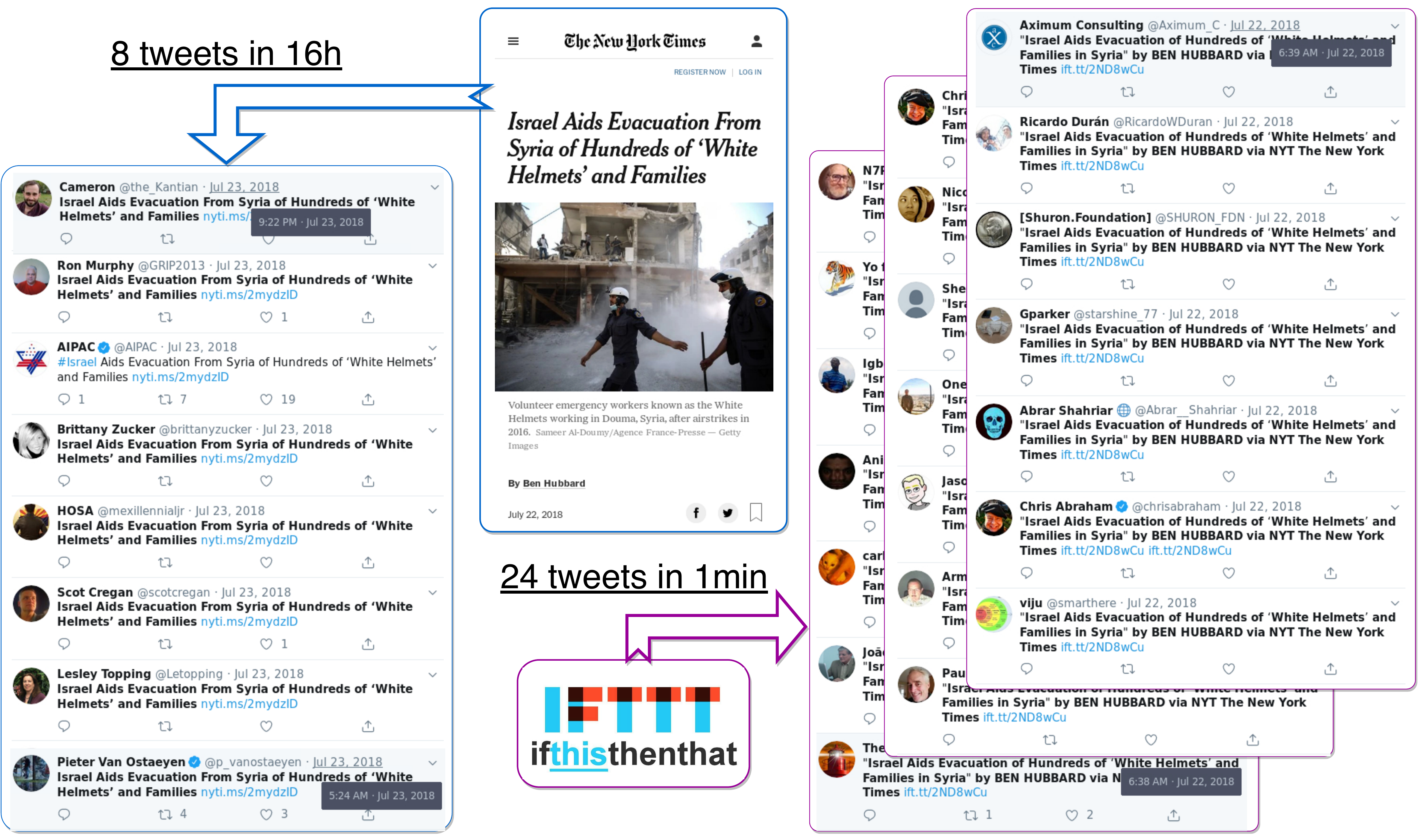}
    \caption{Organic \textit{versus} automated news sharing. On the left, tweets shared by accounts using the ``share button'' at \url{nytimes.com}, and on the right, tweets generated via the IFTTT automation service. The latter produces many more tweets in a short interval, and is detected in our coordination network. On the other hand, despite its popularity, \emph{The New York Times} does not meet the timing constraints of our methods.}
    \label{fig:ny_example}
\end{figure*}

Fig.~\ref{fig:ny_example} illustrates an example of the problem described above using one of the most recurrent headlines in our dataset (``Israel aids evacuation of hundreds of `White Helmets' and families''). 
Searching for the text generated automatically by the share button of \textit{The New York Times} website results in eight tweets total, spanning more than 16 hours: all were created on July 23, 2018 but no two in the same minute. 
On the other hand, searching for the text shared by a set of accounts using IFTTT, identified by our method, yields three times as many tweets within a one-minute interval, and one day earlier than the first share via \url{nytimes.com}. Moreover, even if these two sets of tweets happened at the same time, the similarity score of the two texts is 0.65. Thus, given the similarity threshold used in this study, these accounts would be aggregated into separated groups.

Lastly, it is outside the scope of this paper to unveil the motivations of the actors behind these coordinated groups. Some might be trying to push certain narratives pro- or anti-White Helmets due to ideology, or because they have been paid. Others may want to exploit the visibility of the subject to get profitable traffic. Regardless of their intentions, these groups are manipulating online discourse. 


\section{Conclusions}

In this paper we explored different aspects of inauthentic promotion of content and/or accounts on Twitter in the context of the White Helmets disinformation campaign.

First, we used a simple retweet approach trying to detect automatic retweets. We were able to identify groups of \textit{promoter} and \textit{promoted} accounts, even though their coordination cannot be proven.

Second, we explored text similarity. Most quasi-identical tweets are created in the first 10 seconds after the original content is posted. This suggests these copies are more likely to be the result of automation than a natural adoption and transmission of information. 
We identified different strategies used by the coordinated groups, ranging from just a few accounts repeatedly sharing the exact same content to dozens of accounts synchronously posting the same content in allegedly distinct media outlets.

The framework used here is sensitive to parameters, especially the similarity threshold. We showed that being more rigorous to avoid false positives leads to fewer and smaller groups, as expected. However, fine tuning of the threshold enables the detection of complex components with non-trivial structure.
More work is needed to formulate formal statistical methods to define suspicious connections based on comparisons with appropriate null models rather than strict filters.

We hope the approaches proposed here can be used by social media platforms, researchers, and journalists to promote fair and reliable debates online. Ultimately, the early detection of disinformation campaigns aims to reduce their damage to the general public.

\subsection*{Acknowledgments}
We gratefully acknowledge support from the Knight Foundation, Craig Newmark Philanthropies, and DARPA (contract W911NF-17-C-0094).

\bibliographystyle{ACM-Reference-Format}
\bibliography{references}

\end{document}